\documentclass[11pt]{article}
\usepackage{times}
\usepackage{graphicx}
\usepackage{mathtools}
\usepackage{color}

\renewcommand{\epsilon}{\varepsilon}

\title{\sf 
Dynamics of osmosis in a porous medium
}

\author{Silvana S. S. Cardoso$^*$ \& Julyan H. E. Cartwright$^\dag$
 \\ \\
$^*$ Department of Chemical Engineering and Biotechnology, \\ University of Cambridge,  
Cambridge CB2 3RA, UK \\ \\
$^\dag$  Instituto Andaluz de Ciencias de la Tierra, \\ CSIC--Universidad de Granada, \\
Campus Fuentenueva, E-18071 Granada, Spain.}

\date{Version of \today}

\begin{document}

\maketitle


\vspace{0.5cm}

\emph{
We derive from kinetic theory, fluid mechanics, and thermodynamics the minimal continuum-level equations governing the flow of a  
binary, non-electrolytic mixture in an isotropic porous medium with osmotic effects.  For dilute mixtures, these equations are linear and in 
this limit provide a theoretical basis for the widely-used semi-empirical relations of Kedem and Katchalsky (1958), which have hitherto 
been validated experimentally but not theoretically.  The above  linearity between the fluxes and the driving forces breaks down for concentrated or non-ideal mixtures, for which our equations go beyond the Kedem--Katchalsky formulation.
We show that the heretofore empirical solute permeability coefficient reflects the momentum transfer 
between the solute molecules that are rejected at a pore entrance and the solvent molecules entering the pore space; it can be related to the inefficiency of a Maxwellian demi-demon.}

\section{Introduction}

It seems that there is currently no correct theoretical development of the fundamental 
equations describing the physics of transport in a porous medium with osmotic effects. 
At present, all work to model osmotic flow in a porous medium at the continuum level ultimately derives from the semi-empirical 1958 
formulation of Kedem and Katchalsky \cite{kedem}. 
According to Kedem--Katchalsky,
for dilute solutions,  the molar flux of solute (species 1), $N_1$, and volume flux of solvent (species 2), $u_2$, across a membrane permeable to the 
solvent but only partially permeable to the solute, are
\begin{eqnarray}
 u_2&=&L \left( -\delta p+\sigma RT \delta c_1 \right), \label{KK} \\
N_1&=&-wRT \delta c_1 + \left( 1-\sigma \right) u_2 c_1. \nonumber
\end{eqnarray}
These relations were obtained from non-equilibrium thermodynamics under the assumption of linearity between the fluxes and the driving 
forces.   Here $R$ is the universal gas constant and the  
temperature $T$ is assumed constant; $\delta p$ and $\delta c_1$ are the pressure and molar concentration differences across the membrane, and $L$ is a transport 
coefficient. The reflection coefficient $\sigma$ measures the fraction of solute molecules that are reflected by the membrane 
\cite{staverman}, taking the value of one for a perfectly semipermeable membrane and zero for a completely permeable one. The 
solute permeability coefficient $w$ is null for a semipermeable membrane ($\sigma=1$);  little is known about the physical meaning of 
$w$.   The phenomenological coefficients $w$, $\sigma$, and $L$ are measured, for a given solute and membrane, by carefully 
designed experiments \cite{medved}.  Post Kedem and Katchalsky, earlier theoretical proposals for the interaction of osmosis and viscous flow in a porous medium
 include the use of a potential-energy field \cite{manning} and of friction factors \cite{richardson}, but neither of these approaches relate the ad-hoc coefficients introduced therein to the properties of the solution and of the porous medium. Later work relied on a dusty-gas type model \cite{mason, mason2}, but has 
been proven erroneous \cite{kerkhof} owing to a double count of the viscous forces in the fluid and at the fluid--solid boundaries; this 
model also emphasized the significance of a `partial osmotic pressure' but failure to distinguish between equilibrium static and non-equilibrium flow situations may have caused confusion. A similar error has propagated into subsequent literature \cite{biesheuvel}.  More recent models \cite{revil,revil2} have taken into account electrostatic effects outside the pore entrance and exit to describe the osmotic pressure in terms of an electric double-layer potential; however, these works do not apply to a non-electrolytic system.
In this work, we perform a momentum balance at the molecular level to derive the minimal continuum-level equations for the flow of a binary, non-electrolytic 
mixture in a porous medium in the presence of osmotic effects. We discuss the conditions under which these equations may be reduced 
to the simplified, semi-empirical form above, and we address the physical significance of the solute permeability coefficient, $w$. 

The classical treatment of osmosis considers a system in thermodynamic equilibrium.  While the early works of van't Hoff \cite{van'thoff} 
and Rayleigh \cite{rayleigh} explained osmosis in terms of the work done by the rebounding molecules of solute on a selective 
(semipermeable) membrane, the same phenomenon was later described by Gibbs onwards in terms of the free energy and chemical 
potential \cite{gibbs}. Different disciplines have preferred one or the other of these approaches to derive the classical thermostatic result, but 
the kinetic and thermodynamic theoretical treatments are entirely equivalent \cite{lachish}. However, to quantify the evolution of a system 
towards such equilibrium, the flux laws governing the flows of solute and solvent are necessary.  Recent osmotic research has focussed 
on molecular-dynamics simulations of flow, for example in nanopores \cite{molecular1}, in nanotube arrays \cite{molecular2}, and in 
nanofluidic diodes \cite{picallo}. But, in spite of such numerical studies, little is known about the key intermolecular and molecule--pore 
interactions that drive osmotic flow in a pore and how these relate to continuum-level properties of the fluid and the porous matrix. Yet, it 
is this translation of the molecular behaviour to a mesoscale involving many pore lengths, connecting the atomic scale and the macroscopic scale, that is of utmost importance for the  understanding of the role of osmosis in all its manifold applications in physics, chemistry, and biology. Thus we follow in the spirit of Einstein's study of Brownian motion \cite{einstein} in coupling a kinematic approach to osmosis with fluid mechanics.

In order to develop a simple theoretical argument, we focus on core mechanisms and make the following assumptions:
(i) The porous medium is rigid, isotropic and homogeneous.
(ii) The porous medium and the fluid are in thermal equlibrium and isothermal.
(iii) The mixture is binary and non-electrolytic.
(iv) The solute is inviscid and the solvent is viscous.  The introduction of solute-solute interactions through a viscosity for the solute is straightforward \cite{kerkhof}, but complicates the mathematical presentation. We have opted to keep the model as simple as possible.
(v) The interactions between solute and solvent molecules are represented by the 
Maxwell--Stefan diffusivity of the solute in a binary mixture of solute and the solvent.  A discussion of this assumption is provided in the text.
(vi) We neglect possible chemical reaction of the species with each other and solvation at the pore wall.  These effects may be introduced in a further development of the model \cite{biesheuvel}.
(vii) We assume the flow has low Reynolds and P\'eclet numbers.  The validity of this assumption is discussed in the text.

\begin{figure}[tb]
\begin{center}
\includegraphics[width=\columnwidth]{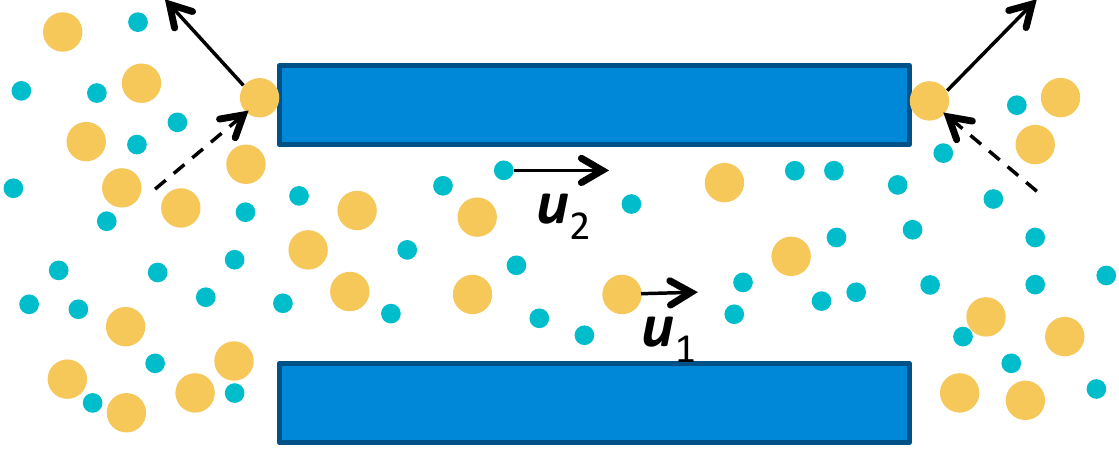}  
\end{center}
\caption{Flow of solute and solvent molecules near a pore entrance and exit.  Some of the solute molecules rebound from the pore 
entrance and subsequently transfer part of their momentum to neighbouring solvent molecules through collisions.  A similar process 
happens at the pore exit.  A difference in the concentrations of solute between the entrance and exit creates an osmotic force. (The instantaneous velocity of a single solute molecule impacting at the pore boundary is much larger than the solute average velocity shown inside the pore; the arrows are not drawn to scale.)
}
\label{sketch}
\end{figure}

\section{Derivation}

Consider the isothermal flow of a binary fluid, comprising a non-electrolytic solute species 1 and a solvent species 2, in a porous medium 
(Fig.~\ref{sketch}).  
The medium is completely permeable to the molecules of  a viscous solvent but only partially permeable to an inviscid solute owing to 
chemical or physical effects. Thus, as the solution flows in a given direction within the pore space, a fraction $\sigma$ of the 
molecules of solute are reflected backward after elastic collision with the solid wall at the entrance to the pore; the remainder flow forward into the pore space, where they undergo further elastic collisions. 
The molecules of solvent do not rebound upon striking the solid walls, but stick to the wall and later leave it with zero average velocity parallel to the pore wall \cite{chapman}. The mass and energy fluxes of incident and emitted molecules are equal.  However,  momentum is not equal for the fluxes of incident and released molecules at the wall; indeed, viscous shear will transfer momentum to the pore wall. We next quantify each of the momentum changes undergone by the solute and the solvent molecules. As the molecules of solute move through the 
pores of the solid matrix, they change momentum owing to two different types of interactions: collisions with molecules of solvent within 
the pore and collisions with the walls of the solid matrix.  The rate of change of momentum of molecules of species 1, per unit volume of mixture 
in a pure fluid medium, resulting from collision with molecules of species 2 is $RT c_1 c_2 \left(\boldsymbol{u_1}-
\boldsymbol{u_2} \right) / ((c_1+c_2) D_{12})$ \cite{taylor-krishna}.  In elementary kinetic theory of diffusion 
\cite{chapman}, the product of concentrations of solute and solvent $c_1 c_2$ reflects the number of collisions and the difference in the 
average velocities $\left(\boldsymbol{u_1}-\boldsymbol{u_2} \right)$, the average momentum exchanged in a single elastic collision of 
smooth, rigid, spherical molecules. Of course, in reality more complex effects may arise  through non-elastic collisions, 
possible multiple molecular encounters, the effect of non-uniformities in composition and pressure on the Maxwellian velocity distribution 
of the molecules, and the presence of internal as well as translational molecular energy \cite{chapman}. However, the physical interpretation of the 
Maxwell--Stefan diffusivity of the solute in the binary mixture of solute and solvent, $D_{12}$, as an inverse drag coefficient remains valid \cite{taylor-krishna}, 
whether the frictional drag exerted by one set of molecules moving through the other arises purely from binary elastic intermolecular 
collisions or from more complex interactions.   
In the porous medium, only a fraction $\left( 1-\sigma \right) $ of solute molecules enters a pore, so that the number of collisions is 
proportionally reduced. Also, the molecules move in tortuous paths around the solid, so that the flux of momentum in any 
particular direction is reduced by a factor $1 / \tau =\overline {\cos^2 \theta}$, where $\theta$ is the inclination of a pore relative to the 
direction specified and the bar represents an average over all pore directions; $\tau$ is the tortuosity of the porous matrix \cite{phillips}.  
The rate of change of momentum of molecules of species 1, per unit volume of mixture in a porous medium, resulting from collisions with 
molecules of species 2 is then
\begin{equation}
  RT\left( 1-\sigma \right) \frac{c_1 c_2 \tau \left(\boldsymbol{u_2}-\boldsymbol{u_1} \right)}{  (c_1+c_2) D_{12}}.
\label{collision_fraction}\end{equation}

The fraction $\sigma$ of the molecules of solute that impact on the solid and rebound at the entrance and exit of a pore undergo a change in momentum, per unit volume 
of fluid, of magnitude
\begin{equation}
 {1 \over 2} \sigma R T \boldsymbol{\nabla} c_1 - \left(-{1 \over 2}\sigma R T \boldsymbol{\nabla} c_1 \right) = \sigma R T \boldsymbol{\nabla} c_1 ,
\label{rebound_fraction}\end{equation}
where $   \sigma R T \boldsymbol{\nabla} c/2 $ is the momentum of molecules leaving the solid surface and $ - \sigma R T \boldsymbol{\nabla} c /2$ is the momentum of the molecules impacting on the solid. This can be viewed as a gradient of osmotic pressure, and $\sigma$ can be related to the ratio of the solute molecule size and the pore size; the solvent molecules are regarded as essentially infinitesimal in size. After the solute molecules rebound from the porous medium, a fraction of their 
momentum is transferred to neighbouring solvent molecules through collisions. This fraction of momentum depends on the distribution of solute and solvent 
molecules near the rebounding surface, which is unknown; we assume it takes a constant average value of $2 \beta$. Thus, 
the change in momentum of the solute molecules caused by collision with the solid surface and with neighbouring solvent molecules, per unit volume of fluid, is \begin{equation}
{1 \over 2} \sigma R T \boldsymbol{\nabla} c_1 +(1-2 \beta) {1 \over 2}\sigma R T \boldsymbol{\nabla} c_1 =(1-\beta )
\sigma R T \boldsymbol{\nabla} c_1. 
\label{demonescape_fraction}\end{equation}
 To the authors' knowledge, the introduction of the parameter $\beta$ representing the exchange of momentum through collision between rebounding solute molecules and solvent molecules near a pore entrance and exit is novel. From a physical point of view, it is clear that such transfer takes place, but how important is it?  We shall confirm later that $\beta$ is indeed non-zero, and can be related to the solute permeability coefficient $w$ used in the phenomenological model of Kedem and Katchalsky (1958) \cite{kedem} and more recently measured experimentally \cite{medved}.

The total change in momentum of the molecules of solute is balanced by the driving force from the gradient in chemical potential of the 
solute, $g_1$, expressed per unit volume of fluid as  \cite{taylor-krishna}
\begin{equation}
  c_1  \boldsymbol{\nabla}_T g_1 = c_1  \boldsymbol{\nabla}_{T,p} g_1+\phi_1 \boldsymbol{\nabla} p,
\label{entropy}
\end{equation}
where $\phi_1$ the volume fraction of the solute in the mixture. In effect this is the driving force for entropy production owing to irreversible processes \cite{sandler, gibbs, fermi}. The momentum balance for the 
solute may thus be written as
\begin{equation}
\begin{multlined}
 c_1  \boldsymbol{\nabla}_{T,p} g_1+\phi_1 \boldsymbol{\nabla} p= 
RT\left( 1-\sigma \right) \frac{c_1 c_2 \tau \left(\boldsymbol{u_2}-\boldsymbol{u_1} \right)}{  (c_1+c_2) D_{12}}
 \\ +\left( 1-\beta \right) \sigma R T \boldsymbol{\nabla} c_1 .
\end{multlined}
\label{solute_momentum_balance}
\end{equation}
Each of the quantities on the righthand side of this equation arise from terms ~(\ref{collision_fraction}) and (\ref{demonescape_fraction}) discussed earlier. A momentum balance for the molecules of solvent leads to a similar equation,
\begin{equation}
\begin{multlined}
 c_2  \boldsymbol{\nabla}_{T,p} g_2+\phi_2 \boldsymbol{\nabla} p=  \\
RT\left( 1-\sigma \right) \frac{c_1 c_2 \tau \left(\boldsymbol{u_1}-\boldsymbol{u_2} \right)}{  (c_1+c_2) D_{12}}  +\beta  \sigma R T \boldsymbol{\nabla} c_1- \frac{\mu_2 }{k_2}  \phi_2 
\boldsymbol{u_2} .
\end{multlined}
\label{solvent_momentum_balance}
\end{equation}
Here $\mu_2$ is the viscosity and $\phi_2$ the volume fraction of the solvent in the mixture; in general, the permeability of the medium to the solvent in 
the presence of the solute, $k_2$, varies with the composition of the mixture. The first term on the right-hand side accounts for the momentum change of the solvent molecules upon collision with the solute molecules in the pore.  The second term quantifies the momentum change of the solvent molecules through collision with rebounding solute molecules at the entrance and exit of the pore; as mentioned above, a fraction $2 \beta $ of the momentum of the rebounding solute molecules is transferred to the solvent. The last term accounts for the loss of momentum of the solvent molecules upon collision and sticking at the solid surface inside the pore, as described earlier.  It quantifies the effect of viscous forces averaged over many 
pore orientations and gives rise to Darcy's law for flow in a porous medium \cite{phillips}.
 
In the momentum balances above for the solute and solvent, we have assumed that the mean free path of the 
molecules is sufficiently smaller than the pore diameter in the solid matrix, so that the fluid may be modelled as a continuum with constant 
transport properties such as viscosity and diffusivity. We have also assumed that the momentum change arising from the acceleration or deceleration of the fluid as it moves 
through the tortuous paths in the porous medium is negligible; this assumption is valid \cite{phillips} for low-Reynolds-number flows such 
that $Re_2=\rho_2 \left|  \boldsymbol{u_2} \right| \delta / \mu_2 \ll 1$, where $\rho_2$ is the density of density of the solvent and $\delta$ 
is the typical pore length scale. The effects of dispersion of a component arising from such tortuous motion has been neglected, which is acceptable when the P\'eclet number of the flow is small, $Pe_i=\left|  \boldsymbol{u_i} \right| \delta / D_{12} \ll 1, 
(i=1,2)$ \cite{phillips}.  

The sum of Eqns~(\ref{solute_momentum_balance}) and (\ref{solvent_momentum_balance}) quantifies the pressure gradient in terms of the velocity of the solvent and the osmotic effect of the solute,
\begin{equation}
  \boldsymbol{\nabla} p=  - \frac{\mu_2 }{k_2}  \phi_2 \boldsymbol{u_2}   +\sigma R T 
\boldsymbol{\nabla} c_1 .
\label{nabla_p}\end{equation}
Substituting Eqn.~(\ref{nabla_p}) into (\ref{solute_momentum_balance}) leads to a relation between the velocities of the solute and 
solvent,
\begin{equation}
\begin{multlined}
 \left[ RT\left( 1-\sigma \right) \frac{c_1 c_2 \tau  }{  (c_1+c_2) D_{12}}  \right] \left( \boldsymbol{u_1} - \boldsymbol{u_2} \right)
+ \phi_1 \phi_2 \frac{\mu_2 }{k_2} \boldsymbol{u_2}= \\
- RT \left[ \Gamma_1- \left( \phi_2 -\beta \right) \sigma \right] \boldsymbol{\nabla} c_1 ,
\label{velocity_relation}\end{multlined}
\end{equation}
where we have used $c_i  \boldsymbol{\nabla}_{T,p} g_i=RT \Gamma_i \boldsymbol{\nabla} c_i$ with $\Gamma_i=\left( c_1+c_2- c_i \right)/ 
\left[(c_1+c_2) \left( 1-\phi_i \right) \right] $, valid for an ideal solution; for non-ideal behaviour one may introduce activity coefficients \cite{sandler} 
in a straightforward manner. Solving Eqns~(\ref{nabla_p})  
and (\ref{velocity_relation}) for the velocities of the solute and solvent leads to
\begin{eqnarray}
  \boldsymbol{u_1} &=&{(c_1+c_2)D_{12} \over {\tau c_1 c_2}} \left[\frac{\phi_1}{RT(1-\sigma)}\boldsymbol{\nabla} p-\frac{\Gamma-(\phi_2-\phi_1-\beta) \sigma}{1-\sigma}   \boldsymbol{\nabla} c_1 \right]+\boldsymbol{u_2} , 
\nonumber \\
  \boldsymbol{u_2} &=& \frac{k_2}{\mu_2 \phi_2}  \left( -\boldsymbol{\nabla} p   +\sigma R T  \boldsymbol{\nabla} c_1 \right) .
\label{full_relation}\end{eqnarray}
The volumetric flux of solvent and the molar flux of solute, per unit area of porous medium, are respectively given by
\begin{eqnarray}
  \boldsymbol{u} &=& \epsilon \boldsymbol{u_2} =  \frac{k_2 \epsilon}{\mu_2 \phi_2 }  \left( -\boldsymbol{\nabla} p   +\sigma R T  
\boldsymbol{\nabla} c_1 \right) ,\label{full_eqns2} \\
  \boldsymbol{N_1}&=& \left(1 -\sigma \right) c_1 \epsilon \boldsymbol{u_1} = \nonumber \\
&&{c_1+c_2 \over c_2}{\epsilon  D_{12}  \over \tau} \left( \frac{\phi_1}{RT} \boldsymbol{\nabla} p-\left[ \Gamma-(\phi_2-\phi_1-\beta) \sigma \right]  \boldsymbol{\nabla} c_1 \right)+ \left( 1-\sigma \right) c_1 \boldsymbol{u} ,
\nonumber
\end{eqnarray}
where $\epsilon$ is the porosity of the medium. We recognize in $\epsilon  D_{12}  / \tau$ the effective diffusivity of the solute in the 
solvent in the porous medium.
For a dilute solution (i.e., in the limit $\phi_1 \rightarrow 0$, $\phi_2 \rightarrow 1$),  Eqns~(\ref{full_relation})  
 simplify to
\begin{eqnarray}
  \boldsymbol{u_1} &=&-\frac{1-(1-\beta) \sigma}{1-\sigma}  {D_{12} \over {\tau c_1}} \boldsymbol{\nabla} c_1+\boldsymbol{u_2} , 
\nonumber \\
  \boldsymbol{u_2} &=& \frac{k_2}{\mu_2 }  \left( -\boldsymbol{\nabla} p   +\sigma R T  \boldsymbol{\nabla} c_1 \right) .
\end{eqnarray}
These relations show that the slip velocity between the solute and the solvent arises essentially from the transfer of momentum from 
inter-species molecular collisions, i.e., frictional drag between the species, but not from the presence of the solid matrix. The flow of 
solvent is affected by viscous stresses between the fluid and the solid matrix, and the gradient in osmotic pressure owing to the solute. 
For a dilute solution, the volumetric flux of solvent and the molar flux of solute, per unit area of porous medium, are respectively given by
\begin{eqnarray}
  \boldsymbol{u} &=& \epsilon \boldsymbol{u_2} =  \frac{k_2 \epsilon}{\mu_2 }  \left( -\boldsymbol{\nabla} p   +\sigma R T  
\boldsymbol{\nabla} c_1 \right) ,\label{final_eqns} \\
  \boldsymbol{N_1}&=& \left( 1-\sigma \right) c_1 \epsilon \boldsymbol{u_1} = \nonumber \\
&&- \left[ 1-(1-\beta) \sigma \right] {\epsilon  D_{12}  \over \tau} \boldsymbol{\nabla} c_1+ \left( 1-\sigma \right) c_1 \boldsymbol{u} ,
\nonumber
\end{eqnarray}
 Equations~(\ref{final_eqns}) describe the flow of an ideal dilute binary mixture in an isotropic porous 
medium, with osmotic effects arising from the interaction of the solute molecules with the solid matrix.  The derivation presented here 
may be easily extended, for instance, to a multicomponent mixture with non-ideal behaviour, and to include the gravitational force.

\section{Discussion}

Equations~(\ref{final_eqns}) have the structure of the semi-empirical equations proposed by Kedem and Katchalsky \cite{kedem} given 
in Eqs~(\ref{KK}).  The coefficients are expressed in terms of the properties of the solute, the solvent, and the porous matrix, and satisfy 
Onsager's reciprocal relation \cite{onsager}  in that $\left( \partial  \boldsymbol{u} / {\partial c_1} \right)_p= R T \left[ \partial  \left( \boldsymbol{N_1} / c_1-
\boldsymbol{u} \right) / \partial p \right]_{c_1}$. The present work shows that linearity between the fluxes and the driving forces holds for 
dilute, ideal mixtures, but not for more concentrated or non-ideal ones, for which the slip velocity between the solvent and solute is 
complex (see Eq.~(\ref{velocity_relation})). The solute permeability $w$ reflects the momentum of the solute 
molecules after rebounding near a pore entrance and exit.  It thus represents the efficiency of the sorting process being carried out, and so we might see the coefficient $\beta$  as the inefficiency of the particular Maxwellian demi-demon  of the pore (a complete Maxwell demon \cite{maxwell,maxwell_demon} would require two semipermeable membranes back to back, as Szilard discussed \cite{szilard,leff}).

We hope that this derivation will be of utility to the many people who use the Kedem--Katchalsky equations, which have hitherto 
been validated experimentally but not theoretically. The present work moreover goes beyond Kedem and Katchalsky to the nonlinear regime of concentrated or non-ideal mixtures. We anticipate that it will stimulate future 
molecular-dynamical simulations to explore the role of inter-species momentum transfer at the entrance and exit of nanopores on 
osmosis, and their impact on the continuum-level behaviour of the fluid in a porous medium.

\paragraph{Acknowledgements} J.C. acknowledges the financial support of Spanish 
MICINN grant FIS2010-22322-C02.

\bibliographystyle{unsrt} 
\bibliography{osmosis}

\begin{thebibliography}{10}

\bibitem{kedem}
O.~Kedem and A.~Katchalsky.
\newblock Thermodynamic analysis of the permeability of biological membranes to
  non-electrolytes.
\newblock {\em Biochim. Biophys. Acta}, 27:229, 1958.

\bibitem{staverman}
A.~J. Staverman.
\newblock The theory of measurement of osmotic pressure.
\newblock {\em Rec. Trav. Chim. Pays-Bas}, 70:344, 1951.

\bibitem{medved}
I.~Medved and R.~Cerny.
\newblock Osmosis in porous media: A review of recent studies.
\newblock {\em Microporous Mesoporous Materials}, 170:299, 2013.

\bibitem{manning}
G.~S. Manning.
\newblock Binary diffusion and bulk flow through a potential-energy profile: A
  kinetic basis for the thermodynamic equations of flow through membranes.
\newblock {\em J. Chem. Phys.}, 49:2668, 1968.

\bibitem{richardson}
I.~W. Richardson.
\newblock Some remarks on the {Kedem--Katchalsky} equations for
  non-electrolytes.
\newblock {\em Bull. Math. Biophys.}, 32:237, 1970.

\bibitem{mason}
L.~F. {del Castillo} and E.~A. Mason.
\newblock Generalization of membrane reflection coefficients for non ideal, non
  isothermal, multicomponent systems with external forces and viscous flow.
\newblock {\em J. Membrane Sci.}, 28:229, 1986.

\bibitem{mason2}
E.~A. Mason and H.~K. Lonsdale.
\newblock Statistical-mechanical theory of membranes transport.
\newblock {\em J. Membrane Sci.}, 51:1, 1990.

\bibitem{kerkhof}
P.~J. A.~M. Kerkhof.
\newblock A modified {Maxwell-Stefan} model for transport through inert
  membranes: the binary friction model.
\newblock {\em Chem. Eng. J.}, 64:319, 1996.

\bibitem{biesheuvel}
P.~M. Biesheuvel.
\newblock Statistical-mechanical theory of membranes transport.
\newblock {\em J. Colloid Interface Sci.}, 355:389, 2011.

\bibitem{revil}
A.~Revil and P.~Leroy.
\newblock Constitutive equations for ionic transport in porous shales.
\newblock {\em J. Geophys. Res.}, 109:B03208, 2004.

\bibitem{revil2}
P.~Leroy A.~Revil and K.~Titov.
\newblock Characterization of transport properties of argillaceous sediments:
  Application to the {Callovo-Oxfordian} argillite.
\newblock {\em J. Geophys. Res.}, 110:B06202, 2005.

\bibitem{van'thoff}
J.~H. van't Hoff.
\newblock Die {Rolle} des osmotischen {Druckes} in der {Analogie} zwischen
  {L\"osungen} und {Gasen} (the role of osmotic pressure in the analogy between
  solutions and gases).
\newblock {\em Z. phys. Chem.}, 1:481, 1887.

\bibitem{rayleigh}
{Lord Rayleigh}.
\newblock The theory of solution.
\newblock {\em Nature}, 55:253--254, 1897.

\bibitem{gibbs}
J.~W. Gibbs.
\newblock Semi-permeable films and osmotic pressure.
\newblock {\em Nature}, 55:461, 1897.

\bibitem{lachish}
U.~Lachish.
\newblock Osmosis and thermodynamics.
\newblock {\em Am. J. Phys.}, 75:997, 2007.

\bibitem{molecular1}
A.~V. Raghunathan and N.~R. Aluru.
\newblock Molecular understanding of osmosis in semipermeable membranes.
\newblock {\em Phys. Rev. Lett.}, 97:024501, 2006.

\bibitem{molecular2}
A.~Kalra, S.~Garde, and G.~Hummer.
\newblock Osmotic water transport through carbon nanotube membranes.
\newblock {\em Proc. Natl. Acad. Sci. U.S.A.}, 100:10175, 2003.

\bibitem{picallo}
C.~B. Picallo, S.~Gravelle, L.~Joly, E.~Charlaix, and L.~Bocquet.
\newblock Nanofluidic osmotic diodes: Theory and molecular dynamics
  simulations.
\newblock {\em Phys. Rev. Lett.}, 111:244501, 2013.

\bibitem{einstein}
A.~Einstein.
\newblock {\em Investigations on the Theory of the Brownian Movement}.
\newblock Dover, 1956.

\bibitem{chapman}
S.~Chapman and T.~G. Cowling.
\newblock {\em The Mathematical Theory of Non-Uniform Gases}.
\newblock Cambridge University Press, 3rd edition, 1970.

\bibitem{taylor-krishna}
R.~Taylor and R.~Krishna.
\newblock {\em Multicomponent Mass Transfer}.
\newblock John Wiley and Sons, 1993.

\bibitem{phillips}
O.~M. Phillips.
\newblock {\em Flow and Reactions in Permeable Rocks}.
\newblock Cambridge University Press, 1991.

\bibitem{sandler}
S.~I. Sandler.
\newblock {\em Chemical and Engineering Thermodynamics}.
\newblock John Wiley and Sons, 3rd edition, 1999.

\bibitem{fermi}
E.~Fermi.
\newblock {\em Thermodynamics}.
\newblock Dover, 1956.

\bibitem{onsager}
L.~Onsager.
\newblock Reciprocal relations in irreversible processes. {I}.
\newblock {\em Phys. Rev.}, 37:405, 1931.

\bibitem{maxwell}
J.~C. Maxwell.
\newblock {\em Theory of Heat}.
\newblock (Reprinted, Dover 2001), 1871.

\bibitem{maxwell_demon}
K.~Maruyama, F.~Nori, and V.~Vedral.
\newblock The physics of {Maxwell's} demon and information.
\newblock {\em Rev. Mod. Phys.}, 81:1, 2009.

\bibitem{szilard}
L~Szilard.
\newblock {\"U}ber die {Entropieverminderung} in einem thermodynamischen
  {System} bei {Eingriffen} intelligenter {Wesen} (on the decrease of entropy
  in a thermodynamic system by the intervention of intelligent beings).
\newblock {\em Z. Phys.}, 53:840, 1929.

\bibitem{leff}
H.~S. Leff and A.~F. Rex.
\newblock Entropy of measurement and erasure: {Szilard's} membrane model
  revisited.
\newblock {\em Am. J. Phys.}, 62:994, 1994.

\end{thebibliography}

\end{document}